# Pre-earthquake electromagnetic emissions with critical and tricritical behavior before the recent Durrës (Albania) (Mw=6.4, 26-11-2019) and Chania (Greece) (Mw=6.1, 27-11-2019) earthquakes.

**Stelios M. Potirakis[1,*], Yiannis Contoyiannis[1], Grigorios E. Koulouras[1], Nikolaos S. Melis[2], Konstantinos Eftaxias[3] and Constantinos Nomicos[1]**

[1] Department of Electrical and Electronics Engineering, University of West Attica, Campus 2, 250 Thivon and P. Ralli, Aigaleo, Athens GR-12244, Greece
[2] Institute of Geodynamics, National Observatory of Athens, Lofos Nimfon, Thissio, Athens GR-11810, Greece
[3] Department of Physics, Section of Solid-State Physics, University of Athens, Panepistimiopolis, GR-15784, Zografos, Athens, Greece
[*] Correspondence: spoti@uniwa.gr; Tel.: +30 2105381550; FAX: +30 2105381514

**Abstract**: In this brief report we present evidence for the dynamics of the earthquake (EQ) preparation processes that led to two strong events (Mw>6) that took place in Durrës (Albania) and Chania (West-Crete, Greece) on 26 and 27-11-2019, respectively. Specifically, MHz fracto-electromagnetic emissions (EME) recorded by our telemetric stations presented critical fluctuations prior to each one of these EQs. Subsequently, the EME possibly related to each one of these EQs evolved differently towards the occurrence of the EQ event: the ones possibly related to Chania EQ departed from critical state according to the symmetry breaking phenomenon, while the ones possibly related to Durrës EQ departed from critical state according to the tricritical crossover phenomenon. The analysis was performed by means of the method of critical fluctuations (MCF).

## 1  Introduction

Earthquakes (EQs) are large-scale fracture phenomena in the Earth's heterogeneous crust. Fracture-induced electromagnetic emissions (EME) can be considered as the so-called precursors of general fracture. They are detectable both at a laboratory and a geophysical scale. These precursors allow a real time monitoring of damage evolution during mechanical loading in a wide frequency spectrum ranging from kHz to MHz. Our main observational tool is the monitoring of the fractures which occur in the focal area before the final break-up by recording their kHz-MHz EME. Clear kHz to MHz EM anomalies have been detected over periods ranging from a few days to a few hours to destructive EQs. MHz EM anomalies have been found to be systematically emerging prior to kHz EM anomalies both in laboratory and geophysical scale (Eftaxias and Potirakis, 2013; Eftaxias et al., 2013, 2018 and references therein, Potirakis et al., 2015, 2016, 2019a).

Based on the "Method of Critical Fluctuations" (MCF) (Contoyiannis and Diakonos, 2000, Contoyiannis at al., 2002, 2013), we have shown that fracture induced MHz EMEs recorded prior significant EQs present criticality characteristics , implying that they emerge from a system in critical state (Contoyiannis et al., 2010; Potirakis et al., 2015, 2016).

The emerged "EM critical time window" (CW) can be described in analogy with a phase transition of second order in equilibrium. We note that using the natural time method, which can be successfully applied even if a limited number of data are available, we have shown that seismicity around the epicenter of the impending EQ presents also criticality characteristics (Potirakis et al., 2013, 2015, 2016). This finding indicates that these two different observables, which are measured in the same pre-seismic time period, namely, a few days before the EQ occurrence, could be attributed to the same Earth system, namely, the EQ generation process happening around the fault zone, enhancing the hypothesis for the possible relation of the recorded EME with the subsequent EQ. We have proposed that the observed MHz EM anomaly is due to the fracture of the highly heterogeneous system that surrounds the formation of strong brittle and high-strength entities (asperities) distributed along the rough surfaces of the main fault sustaining the system. We have also proposed that the finally emerged abrupt pulse-like kHz EM anomaly, which shows characteristics of a first order phase transition is due to the fracture of the aforementioned large high-strength entities themselves (Contoyiannis et al, 2015).

In statistical physics, a tricritical point is a point in the phase diagram of a system at which the two basic kinds of phase transition, that is the second order transition and the first order transition, meet. A characteristic property of the area around this point is the co-existence of three phases, specifically, the symmetry area, the low symmetry phase, and an intermediate "mixing" phase. A passage through the area, around the tricritical crossover, from the second order phase transition to the first order transition through the intermediate "mixing" state constitutes a tricritical crossover (TC). Thus, in terms of fracture-induced EME, such a situation would be manifested by an "EM tricritical time window" emerging in the observed preseismic EME time series after the appearance of an "EM critical time window". The appearance of the aforementioned EME tricritical signature indicates that the progress to the first order phase transition of the fracture system, namely, the EQ occurrence, is approaching (Contoyiannis et al, 2015; Potirakis et al., 2016). Another way for departure from critical state in course of the evolution towards the EQ occurrence is the so-called "symmetry breaking" phenomenon (SB) (Contoyiannis et al, 2018; Potirakis et al., 2019b). Note that in the symmetry phase there is a single fixed point (minimum of the Landau free energy $U(\phi)$ vs. the order parameter $\phi$) while in the broken symmetry phase there are two fixed points. Correspondingly, the distribution of the order parameter values (that is the amplitude values of fracture-induced EME) in the first case is of unimodal form (one lobe), while in the second case is of bimodal form (two lobes). This happens because a fixed point attracts a high number of values of the order parameter close to it determining the form of the distribution. If such a situation is observed after a "EM critical time window", the MCF analysis should still detect indications of critical dynamics until the complete departure from critical state (as soon as the two lobes of the bimodal distribution become completely separated).

These two ways for departure from criticality are equivalent. It is noted that in terms of fracture-induced EME the departure from criticality can be detected either by TC, or by SB or by both of them. It is possible that the degree of heterogeneity of the Earth's crust at the area where an EQ is prepared determines which of these three scenarios will dominate in each case.

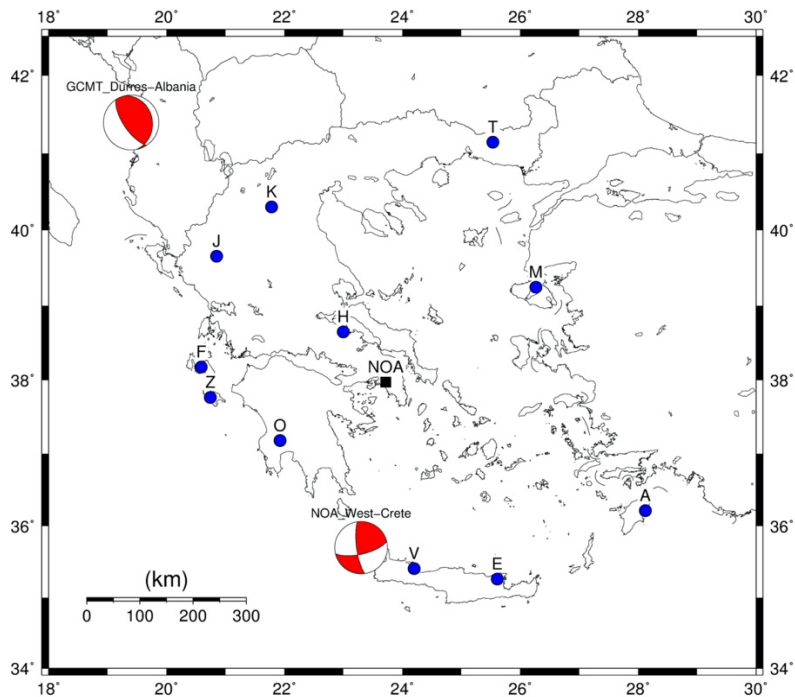

**Fig. 1**. Map showing the 11 fracto-electromagnetic emissions (fracto-EME) stations of the ELSEM-Net (blue circles) in Greece. All of the stations are linked to Institute of Geodynamics of the National Observatory of Athens (NOA-IG) in Athens (black square). Details for the network and the involved instrumentation are available in the online supplementary downloadable material of (Potirakis et al, 2013) and in (Potitakis et al. 2019a). The Durrës and Chania EQs are noted with network and Moment

Tensors (MT): for Durrës from the Global Centroid Moment Tensor project (https://www.globalcmt.org/) and for Chania (West-Crete) from NOA - Institute of Geodynamics (http://bbnet.gein.noa.gr/mt_solution/2019/191127_07_23_43.70_MTsol.html).

Herein, we report we present evidence, in terms of fracture-induced EME, for the dynamics of the earthquake (EQ) preparation processes that led to two recent strong events. The first one occurred 7 km N of Durrës in Albania (41.38º N, 19.42º E) on 26-11-2019, 02:54:18.4 UT, at 23.4 km depth, with a magnitude Mw=6.4. An EM critical time window, immediately followed by an EM tricritical time window were identified ~2.5 days before the EQ in the MHz EME recordings of the Ioannina station (cf. Fig. 1, station code J) of ELSEM-Net (HELlenic Seismo-ElectroMagnetics Network) (http://elsem-net.uniwa.gr), our ground-based fracto-EME station network, spanning across Greece (Fig. 1) (Potirakis et al., 2019a). The second one occurred 79 km W of Chania between the islands of Crete and Antikythera in Greece (35.68º N, 23.26º E) on 27-11-2019, 07:23:43.7 UT, at a depth of 66 km, with a magnitude Mw=6.1. A few hours (~16 h) before, the Ithomi, Mesinia, station (cf. Fig. 1, station code O) of ELSEM-Net recorded MHz EME with critical time window characteristics, immediately followed by MHz EME with symmetry breaking characteristics.

## 2 Data analysis method

The analysis of the recorded data was performed using the method of critical fluctuations (MCF) (Contoyiannis and Diakonos, 2000; Contoyiannis et al., 2002, 2013). Detailed descriptions of all the involved calculations can be found elsewhere (Contoyiannis et al., 2013) and therefore are omitted here for the sake of brevity and focus on the findings. However, a general description of the employed method follows.

MCF was proposed for the analysis of critical fluctuations in the observables of systems that undergo a continuous phase transition (Contoyiannis and Diakonos, 2000; Contoyiannis et al., 2002). It is based on the finding that the fluctuations of the order parameter, that characterizes successive configurations of critical systems at equilibrium, obey a dynamical law of intermittency of an 1D nonlinear map form. The MCF is applied to stationary time windows of statistically adequate length, for which the distribution of the of waiting times (laminar lengths) $l$ of fluctuations in a properly defined laminar region is fitted by a function $f(l) \sim l^{-p_2} e^{-p_3 l}$. The criteria of criticality are $p_2 > 1, p_3 \approx 0$ (Contoyiannis and Diakonos, 2000; Contoyiannis et al., 2002), while for tricriticality are $p_2 < 1, p_3 \approx 0$ (Contoyiannis et al, 2015).

## 3 Analysis results

### 3.1 Durrës EQ

Part of the MHz recordings of the Ioannina station possibly associated with the recent Mw=6.4 Durrës EQ is shown in Fig. 2a. This was recorded ~2.5 days before the occurrence of the EQ. This time-series excerpt, having a total length of 7.5h (27000 samples) was analyzed by the MCF method and was identified to be a CW. As already mentioned in Introduction, CWs are time intervals of the MHz EME signals presenting features analogous to the critical point of a second order phase transition (e.g., Contoyiannis et al, 2015). The main steps of the MCF analysis (e.g., Contoyiannis et al, 2015; Potirakis et al., 2016) on the specific time-series are shown in Fig. 2b- Fig. 2d. As Fig. 2c shows, the obtained plot of the $p_2, p_3$ exponents vs. $\phi_l$ apparently satisfy the criticality conditions, $p_2 > 1$, $p_3 \approx 0$, for a wide range of end points $\phi_l$, revealing the power-law decay feature of the time-series that proves that the system is characterized by intermittent dynamics; in other words, the MHz time-series excerpt of Fig. 2a is indeed a CW.

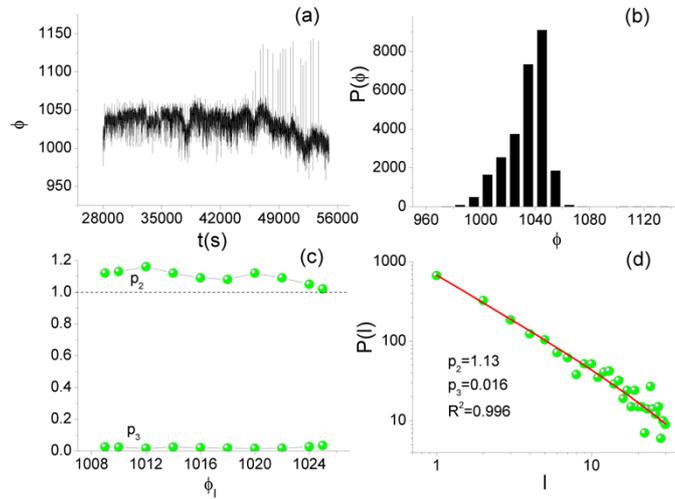

**Fig. 2**. (a) The 27000 samples long critical window of the MHz EME that was recorded before the Durrës EQ. (b) Amplitude distribution of the signal of Fig. 2a. (c) The obtained exponents $p_2, p_3$ exponents vs. different values of the end of laminar region $\phi_l$. The horizontal dashed line indicates the critical limit ($p_2 = 1$). (d) Representative example of the involved fitting for the laminar distribution resulting for a specific end point. The solid line corresponds to the fitted function (cf. to text in Sec. 2) with the values of the corresponding exponents $p_2, p_3$ also noted.

Immediately after the emergence of the CW, part of the MHz recordings of the same station presented characteristics close to tricriticality, as shown by Fig. 3 where the tricriticality conditions $p_2 < 1, p_3 \approx 0$ are satisfied for a wide range of laminar lengths. This EME time series excerpt has a total length of ~2.77h (10000 samples).

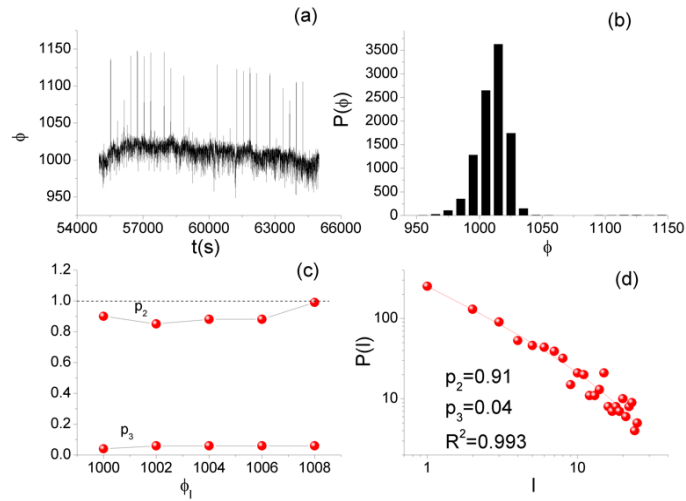

**Fig. 3.** (a) The 10000 samples tricritical window of the MHz EME that was recorded prior to the Durrës EQ. (b) Amplitude distribution of the signal of Fig. 3a. (c) The obtained exponents $p_2, p_3$ exponents vs. different values of the end of laminar region $\phi_l$. The horizontal dashed line indicates the critical limit ($p_2 = 1$). (d) Representative example of the involved fitting for the laminar distribution resulting for a specific end point. The solid line corresponds to the fitted function (cf. to text in Sec. 2) with the values of the corresponding exponents $p_2, p_3$ also noted.

### 3.2 Chania EQ

Part of the MHz recordings of the Ithomi, Mesinia, station possibly associated with the recent Chania (Mw=6.1) EQ is shown in Fig. 4a. This was recorded ~16 h before the occurrence of the EQ. This time-series, having a total length of 10.83 h (39000 samples) was analyzed by the MCF method, revealing three distinct epochs: a critical state, followed by the symmetry breaking phenomenon and then no indication of criticality. Figures 4b-4c show the amplitude value distributions of these three parts of the specific time-series excerpt, while Fig. 5 shows the MCF analysis of the first (Fig. 5a-5b) and the third part (Fig. 5c-5d), respectively.

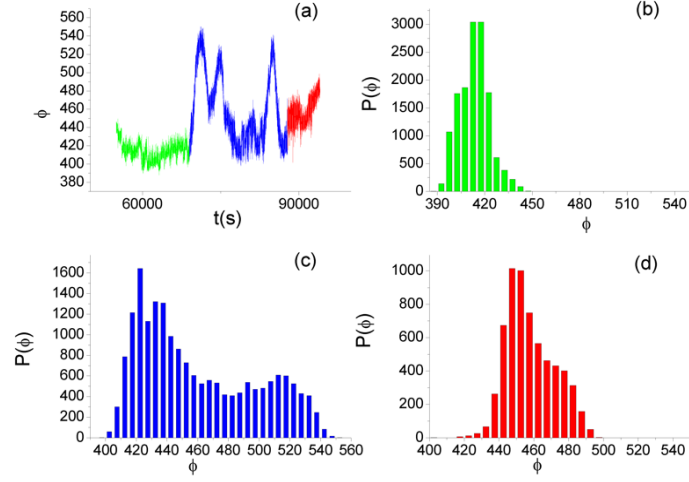

**Fig.4.** (a) The time-series excerpt consists of three distinct epochs. The MCF analysis shows that the green part corresponds to the critical epoch (cf. Fig. 5a-5b), the blue part to the symmetry breaking epoch, and the red part has departed from criticality (cf. Fig. 5c-5d). (b)-(d) Corresponding amplitude values' distributions. The plot of panel (c) demonstrates an important information that is that the right-hand side lobe is smaller than the left-hand side one indicating that the symmetry breaking has almost been completed.

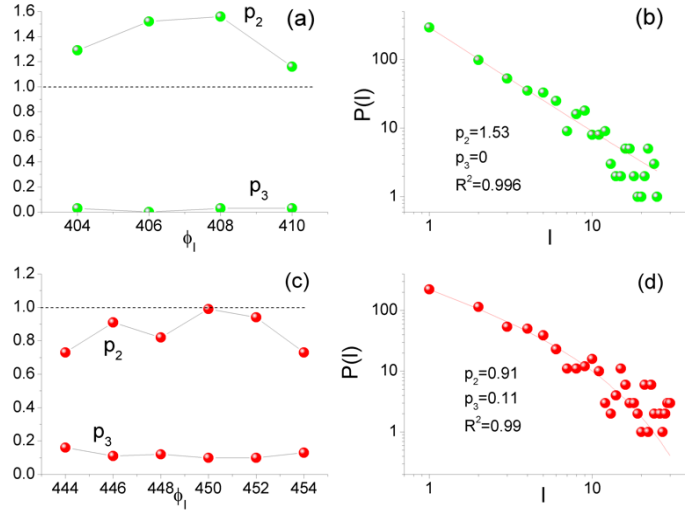

**Fig 5.** (a,c) The obtained exponents $p_2, p_3$ exponents vs. different values of the end of laminar region $\phi_l$. for the critical window (a) and the non-critical epoch (c). The horizontal dashed line indicates the critical limit ($p_2 = 1$). (b, d) Representative examples of the involved fitting for the laminar distribution resulting for a specific end point for the critical window (b) and the non-critical epoch (d). The solid line corresponds to the fitted function (cf. to text in Sec. 2) with the values of the corresponding exponents $p_2, p_3$ also noted.

## 4 Discussion – Conclusion

Based on the method of critical fluctuations, we have shown that the fracture-induced MHz EME recorded prior to the recent significant EQs of Durrës (Albania) and Chania (West-Crete, Greece) occurred on 26 and 27-11-2019, respectively, present critical and tricritical characteristics. The critical behavior of the EME time-series, that was observed first in the timeline for both EQs, implies that this has been generated by a system in critical state. After that, each one followed one of the two equivalent ways for departure from criticality, namely, from the second order phase transition in equilibrium. In the Durrës EQ case, the next appeared behavior was tricritical dynamics in the MHz EME, indicating that the progress to the first order phase transition of the fracture system, namely, the EQ occurrence, is approaching. In the Chania EQ case, the next appeared behavior was symmetry breaking in the MHz EME, indicating that the progress of the second order phase transition was accomplished. Thus, MHz EME possibly related to these two EQs departed from criticality with a different way. It is noted that in the case of the Chania EQ, we observed that symmetry breaking was almost completed (Fig. 4c), and right after a non-critical window appeared but with $p_3$ values not sufficiently far from zero (Fig. 5c). This might be an indication of possible connection between the ways for departure from criticality (symmetry breaking and trictiricality) worth investigating in the future.

**REFERENCES**

Contoyiannis, Y., and F. Diakonos (2000), Criticality and intermittency in the order parameter space, *Phys. Lett. A*, 268, 286 -292, doi: 10.1016/S0375-9601(00)00180-8.

Contoyiannis, Y., F. Diakonos, and A. Malakis (2002), Intermittent dynamics of critical fluctuations, *Phys. Rev. Lett.*, 89, 035701, doi: 10.1103/PhysRevLett.89.035701.

Contoyiannis, Y.F., C. Nomicos, J. Kopanas, G. Antonopoulos, L. Contoyianni, K. Eftaxias (2010), Critical features in electromagnetic anomalies detected prior to the L'Aquila earthquake, *Physica A*, 389, 499-508, doi: 10.1016/j.physa.2009.09.046.

Contoyiannis, Y. F., S. M. Potirakis, and K. Eftaxias (2013), The Earth as a living planet: human-type diseases in the earthquake preparation process, *Nat. Hazards Earth Syst. Sci.*, 13, 125–139, doi: 10.5194/nhess-13-125-2013.

Contoyiannis, Y., S.M. Potirakis, K. Eftaxias, L. Contoyianni (2015), Tricritical crossover in earthquake preparation by analyzingpreseismic electromagnetic emissions, *Journal of Geodynamics*, 84, 40-54, doi: 10.1016/j.jog.2014.09.015.

Contoyiannis, Y., S. M. Potirakis (2018), Signatures of the symmetry breaking phenomenon in pre-seismic electromagnetic emissions, *J. Stat. Mech.* 2018, 083208, https://doi.org/10.1088/1742-5468/aad6ba.

Eftaxias, K., S. M. Potirakis,and T. Chelidze (2013), On the puzzling feature of the silence of precursory electromagnetic emissions, *Nat. Hazards Earth Syst. Sci.*, 13, 2381-2397, doi: 10.5194/nhess-13-2381-2013.

Eftaxias, K., and S. M. Potirakis (2013), Current challenges for pre-earthquake electromagnetic emissions: shedding light from micro-scale plastic flow, granular packings, phase transitions and self-affinity notion of fracture process, *Nonlin. Processes Geophys.*, 20, 771–792, doi:10.5194/npg-20-771-2013.

Eftaxias, K., Potirakis, S.M., Contoyiannis, Y. (2018), Four-stage model of earthquake generation in terms of fracture-induced electromagnetic emissions. In *Complexity of Seismic Time Series: Measurement and Application*; Chelidze, T., Vallianatos, F., Telesca, L., Eds.; Elsevier, Oxford, 2018; pp. 437-502. https://doi.org/10.1016/B978-0-12-813138-1.00013-4

Potirakis, S. M., A. Karadimitrakis, K. Eftaxias (2013), Natural time analysis of critical phenomena: The case of pre-fracture electromagnetic emissions, *Chaos*, 23 (2), 023117(1-14), doi: 10.1063/1.4807908.

Potirakis, S. M., Y. Contoyiannis, K. Eftaxias, G. Koulouras, C. Nomicos (2015), Recent Field Observations Indicating an Earth System in Critical Condition Before the Occurrence of a Significant Earthquake, *IEEE Geoscience and Remote Sensing Letters*, 12(3), 631-635, doi: 10.1109/LGRS.2014.2354374.

Potirakis, S. M., Y. Contoyiannis, N. S. Melis, J. Kopanas, G. Antonopoulos, G. Balasis, C. Kontoes, C. Nomicos, K. Eftaxias (2016), Recent seismic activity at Cephalonia (Greece): a study through candidate electromagnetic precursors in terms of non-linear dynamics, *Nonlin. Processes Geophys.*, 23, 223-240, doi: 10.5194/npg-23-223-2016.

Potirakis, S. M., A. Schekotov, Y. Contoyiannis, G. Balasis, G. E. Koulouras, N. S. Melis, A. Z. Boutsi, M. Hayakawa, K. Eftaxias, C. Nomicos (2019a), On Possible Electromagnetic Precursors to a Significant Earthquake (Mw= 6.3) Occurred in Lesvos (Greece) on 12 June 2017, *Entropy*, 21(3), 241; https://doi.org/10.3390/e21030241

Potirakis, S. M., Y. Contoyiannis, A. Schekotov, T. Asano, M. Hayakawa (2019b), Critical states in the ULF magnetic field fluctuations recorded in Japan prior to the 2016 Kumamoto earthquakes. *Physica A*, 514, 563-572, https://doi.org/10.1016/j.physa.2018.09.070.